\documentclass[12pt, draftclsnofoot, onecolumn]{IEEEtran}

\def\BibTeX{{\rm B\kern-.05em{\sc i\kern-.025em b}\kern-.08em
    T\kern-.1667em\lower.7ex\hbox{E}\kern-.125emX}}
\usepackage{graphicx}
\usepackage[rightcaption]{sidecap}

\begin{document}

\title{Towards Green and Infinite Capacity in Wireless Communication Networks: Beyond The Shannon Theorem}
 
\author{Mohammed S. Elmusrati,\thanks{M. Elmusrati is Full professor and head of communication and systems engineering group at University of Vaasa - Finland. He is also professor at Electrical Engineering department, Benghazi University - Libya.  E-mail: mohammed.elmusrati@uva.fi. All copyrights belong to the author}. }


\maketitle

\begin{abstract}
New and novel way for resources allocation in wireless communication has been proposed in this paper. Under this new method, it has been shown that the required power budget becomes independent of the  number of served terminals in the downlink. However, the required power depends only of the coverage area, i.e. the channel losses at the cell boarder. Therefore, huge number (theoretically any number) of terminals could be supported concurrently at finite and small downlink power budget. This could be very useful to support the downlink signalling channels in HSPA+, LTE, and 5G. It can be very useful also to support huge D2D communication downlinks. Moreover, and based on the same concept, a new system configuration for a single link point-to-point communication has been presented. With this new configuration, the achieved data rate becomes independent of the required transmit power.  This means that any data rate can be achieved at the target BER and with small and finite transmit power. This seems violating with some major results of the Shannon theorem. This issue will be discussed in details in this article.        
\end{abstract}


\section{Introduction}

\PARstart{T}{he} number of devices connected through wireless Internet is estimated to exceed 50 billions by 2020. About $82\%$ of those devices is expected to be driven by wireless networks \cite{ABI}. These devices have vast QoS requirements. Some devices require high data rate (broadband) but with quite flexible latency, some others must have a very small latency but require low data rate and so on. This variations of the required QoS depend on the application (VoIP, Internet surfing, Automation, Monitoring, etc.). In modern wireless broadband networks (e.g., HSPA+, LTE, and 5G), there are many terminals served by the access point (AP). The AP schedules between the those terminals over different dimensions (e.g., time and frequency) using different protocols (e.g, opportunistic scheduling). The APs manage between terminals in the downlink through signalling (or control) channels. Signalling channels have  different tasks such as scheduling grant, HARQ notifications, resources allocations, uplink channel state information, etc.  Unfortunately, considerable resources budget of the AP is allocated to manage such heavy signalling information in the downlink. In some cases, the power budget used to support signalling data can be larger than the power used for the actual user-data transfer \cite{Sig_pow}. In some scenarios, there can be set of many served terminals in the downlink. Nevertheless, only few terminals are active at a time, which are selected according to certain scheduling algorithm. However, all terminals in the set should receive some signalling information frequently.   This fact seriously limits the capacity of the wireless networks. There are some other applications where it is very desirable to have many terminals connected in the downlink. One example is in the area of device-to-device (D2D) communication where there are many actuators in the downlink waiting for control commands from the network. One example in this scenario is the wireless control of street lighting lambs, traffic lights, smart cars,  and generally in wireless automation applications in smart grids and smart cities.   However, there are several critical challenges toward serving huge terminals in the downlink concurrently due to the limited natural resources in terms of power and bandwidth. Although, the bandwidth problem could be handled by using parallel transmission over spatial processing, for example with massive MIMO, however, the problem of the required power is still a critical problem. In 2020, it is expected to have about 42 billion devices connected wirelessly to different core-networks. If it is assumed that, every device requires on average only $0.1$ Watt in the downlink. This simple supposed scenario shows that more than $4000$ megawatt would be needed just for the transmission stage in the downlink.\\ In this paper, a novel technique is introduced with new yet intuitive concepts, which it could support extremely large terminals (actuators, devices, transmission over signalling  channels in mobile networks, etc.) with a very limited power and within the bandwidth required by a single device and without any time scheduling or buffering. The proposed method is based on the features of broadcasting. In the broadcasting systems any number of terminals (theoretically infinite) could be supported in the downlink as far as the terminals are within the broadcasting area (the geographical area where the signal to interference and noise ratio (SINR) is at least achieving the target value). But, in broadcasting systems, all terminals receive the same information. However, in multiuser communication, each terminal receives its own contents, therefore, we cannot utilize the features of broadcasting in multiuser networks with the traditional techniques. However, with a new looking for resources allocation concepts and using some concepts of digital communication, this problem could be solved.

\section{System Model}

The purpose of this paper is to introduce new idea to handle very large load (theoretically can be infinite) in the downlink of wireless networks with finite and limited power budget. Hence, the level of problem treatment will be kept as simple as possible to make clear concepts.  
Assume $N$ active terminals in the downlink of a single access point (AP). The maximum power available at the AP is $P_{max}$.  Therefore, in conventional systems, the total power, which is allocated for all active terminal must achieve the following constraint: 
\begin{equation}
\sum_{i=1}^N P_i\left(t\right) \leq P_{max}
\end{equation}

This means that all terminals should share certain limited power. The achieved performance in the downlink depends on the instantaneous SINR at each terminal input, which is given by
 
\begin{equation}
\gamma_i\left(t\right)=\frac{P_i\left(t\right)G_i\left(t\right)}{G_i\left(t\right)\sum_{j=1, j\ne i}^N P_j\left(t\right)\alpha_{ij}+\sigma_n^2} 
\end{equation}
Where $G_i\left(t\right)$ is the instantaneous channel gain between the access point and the terminal $i$ at time $t$, $\alpha_{ij}$  represents the orthogonality factor between terminals $i$ and $j$, and it depends on the applied multiple access technique, and $\sigma_n^2$  is the average noise power at terminals. The channel gain  $G_i$ is a random process and hence, the SINR is a random process as well.   

If the average target SINR of each terminal is denoted as $\hat{\gamma}_i^T$, the number of supported terminals $N$ must meet the following two conditions (the time symbol $t$ has been dropped for convenience):

\begin{eqnarray} \label{eq1}
E_{G_i}\left[\frac{P_i G_i}{G_i\sum_{j=1, j\ne i}^N P_j \alpha_{ij}+\sigma_n^2}\right] \geq \hat{\gamma}_i^T, \forall i=1,2,...,N \\ 
\sum_{i=1}^N P_i\left(t\right) \leq P_{max} \nonumber
\end{eqnarray}

From Equations (\ref{eq1}), it is clear that the number of supported terminals is limited by mainly two factors, the multiple access method ($\alpha_{ij}$) and the maximum available power $P_{max}$. In ultimate case with using perfect orthogonal technique such as massive beamforming, we may have parallel orthogonal channels so that ($\alpha_{ij}=0 \, \forall i,j$). However, still the number of terminals will be  very limited due to the available power budget $P_{max}$. In this case, the maximum number of terminals $N$ should achieve the following condition: 

\begin{equation} \label{eq5}
\sigma_n^2\sum_{i=1}^N \frac{\hat{\gamma}_i^T}{\hat{G}_i} \leq P_{max}
\end{equation}

where $\hat{G}_i=E_{G_i}\left[G_i\right]$.

It is clear that the required power budget increases linearly with the number of active terminals without limit. This means simply that infinite terminals would need infinite power budget. Assuming a terminal at 1 km away from the access point with a noise figure $=5 dB$, the target SINR is $15dB$, and bandwidth=$5 MHz$, then the downlink power required to support that terminal only will be around 2 Watts (can be computed by using Equation (\ref{eq5})). Therefore, if we have just $50$ terminals with the same conditions, we would need at least 100 watts of downlink power even with perfect orthogonality. This large requirement of downlink power limits the number of supported terminals in the downlink.   

\section{Broadcasting Based Multi-Content Algorithm (BBMA)}

Conceptually, in digital communication the information is sent through finite set of symbols (e.g., binary, 8-PSK, 16-QAM, .. M-ary). For example in binary system, the transmitter sent only one of two possible symbols. If all terminals in the downlink use binary modulation, then all of them share in those two symbols. The information contents is in the sequence of those binary symbols. In this proposal, the radio resources should be allocated according to the number of symbols (not the number of terminals) and shared between all terminals like broadcasting. This could greatly enhance the load capacity of wireless systems and in the same time greatly reduce the energy consumption (green communication). 
Without loss of generality assuming that all terminals in the downlink use the same modulation level (say M-ary), then simply all terminals should receive from the same symbol set or pool $\in \{S_1, S_2,\dots, S_M\}$. This means simply that there could be some terminals receive the same symbol at certain time (synchronization could be handled by the access point in the downlink). In that case, we should not divide the resources between those terminals, which share the same symbol, instead, that symbol should be broadcast to those terminals with power enough to achieve the required SINR for the worst channel user. Therefore, all other terminals will achieve at least the same SINR or better. To perform clear explanation, assume binary system, i.e.,  $S_i \in \{S_1, S_2\}$. In this case, we have just two symbols (or two constellation points) to be transmitted to all terminals. At certain time $t_k$, the AP should check the terminals, which ones should receive symbol $S_1$ and those, which should receive $S_2$. The AP will allocate  power $P_1$ to symbol $S_1$ and $P_2$ to symbol $S_2$. Since, there are two classes of terminals, $class_1\left(t_k\right)$ contains all terminals that should receive $S_1$ at time $t_k$ and $class_2\left(t_k\right)$ contains all terminals that should receive $S_2$.  It is required to keep orthogonality between both classes, and this could be achieved by spatial processing as will be discussed later.  Since, the concept of the proposed algorithm is new and could cause some confusion,  simple numerical example is given next. 
Assume we have $11$ terminals from $T_1$ to $T_{11}$ in the downlink of a single cell with binary modulation (e.g., BPSK). All terminals are divided between two classes where every terminal can be located in only one class at a time. $Class_1$ contains $n_1$ terminals and $Class_2$ contains $n_2$ terminals, and the total number of terminals is $N=n_1+ n_2=11$. The access point (AP) should check how many terminals in each class should receive symbol $S_1$ (and of course the rest should receive the symbol $S_2$). The allocation of symbols between classes can be static or dynamic. In static allocation, the symbols are permanently assigned to classes. For example, $S_1 \rightarrow Class_1$ and $S_2 \rightarrow Class_2$. In dynamic allocation, at every time $t_k$ the symbols could be varying between classes in order to achieve certain criteria such as minimizing the number of moved terminals from one class to another.  Anyway, the AP rearranges the terminals in order to have each class of terminals with the same received symbol at each symbol-time period.  
Assume at certain symbol-time $t_k$, the terminals were distributed between both classes as shown in Table (I). Table (I) shows also the received symbol of every terminal at the next symbol-time $t_{k+1}$.

\begin{table} [htb] \label{Ta}
\caption{Example for BBMA Process.}
\begin{center}
{\tt
\begin{tabular}{|c||c|c|c|c|c|c|}\hline
$Class_1\left(t_k\right)$&$T_1$&$T_4$&$T_5$&$T_7$&$T_{10}$&$T_{11}$\\\hline\hline
Next symbol ($t_{k+1}$)&$S_2$&$S_1$&$S_2$&$S_2$&$S_1$&$S_2$\\\hline
$Class_2\left(t_k\right)$&$T_2$&$T_3$&$T_6$&$T_8$&$T_{9}$&$-$\\\hline
Next symbol ($t_{k+1}$)&$S_1$&$S_2$&$S_2$&$S_1$&$S_1$&$-$\\\hline
\end{tabular}
}
\end{center}
\end{table}

Looking at both classes in Table (I), it shows in $Class_1$ that there are 4 terminals that should receive $S_2$ and 2 terminals that should receive $S_1$ in the next symbol-time period $t_{k+1}$. Furthermore, in $Class_2$ there are 3 terminals should receive $S_1$ and 2 terminals should receive $S_2$. In dynamic allocation scenario, and in order to minimize the movement process between both classes, $S_2$ is assigned to $Class_1$ and $S_1$ is assigned for $Class_2$. Furthermore, the AP should move $\left(T_4,T_{10}\right)$ from $Class_1$ to $Class_2$ and $\left(T_3,T_6\right)$ are moved from $Class_2$ to $Class_1$. Now the available resources at the AP (e.g., power) are allocated between two classes of terminals, instead of dividing the resources among all terminals. With such configuration, it is possible to support any number of terminals as far as they are within the service area of the AP. It can be also used to reduce the power consumption considerably. 
With the two classes configuration (binary system) discussed here, one should guarantee the target SINR for all terminals in each class. This could be achieved by  allocating the power required by the worst terminal (in terms of the  channel gain) in each class. All other terminals in the class (regardless their number, even infinite) will achieve at least the target SINR or higher. 

\section{Orthogonal Classes Through Spatial Processing}

One of the main requirements of this algorithm is the ability to classify terminals in the downlink according to their received symbol. The terminals are distributed spatially randomly. One possible method is to use massive beamforming to achieve orthogonality between terminals in the different classes \cite{massive}. There are several beamforming algorithms could be used \cite{mass2}. For example massive MIMO could be one a strong candidate for this application. However, in this paper, single antenna terminals have been considered and the usage of null-steering beamforming algorithm is considered. 
 
\begin{figure}[h]
\caption{Beamforming Antennas at the Access Point}
\label{Fig1}
\centering
\includegraphics[width=1\textwidth]{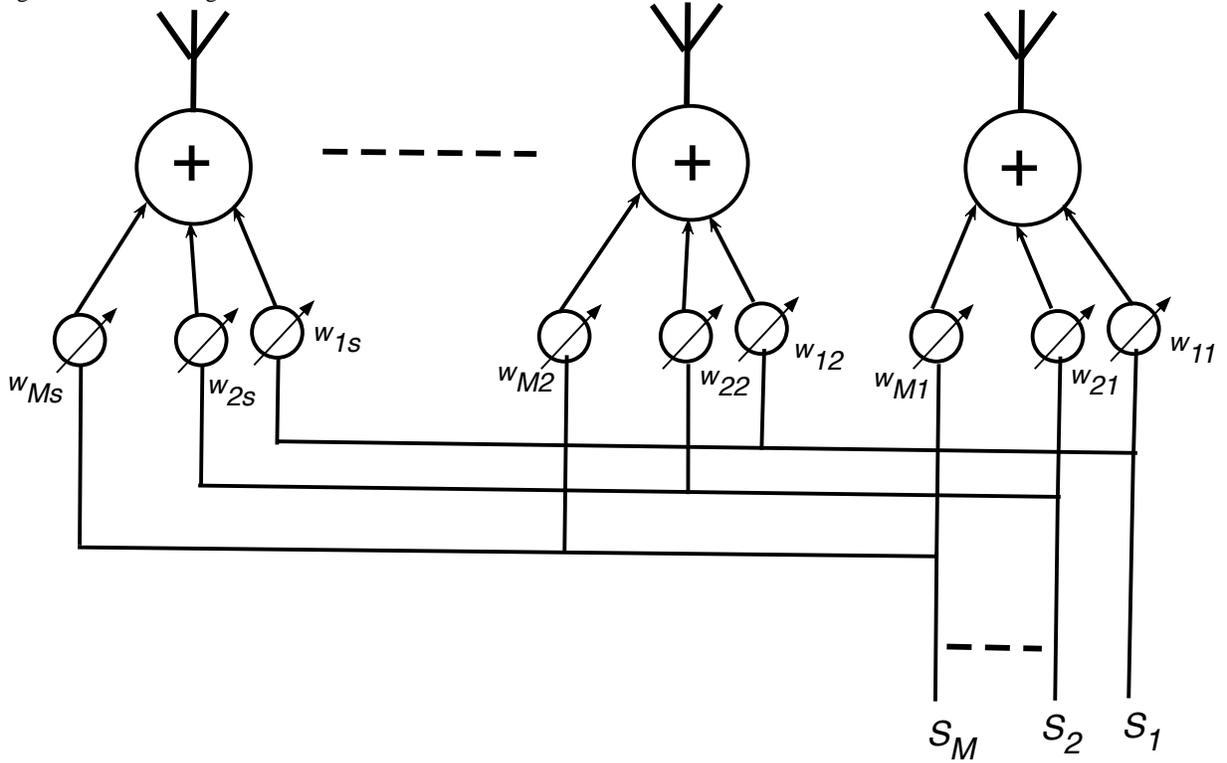}
\end{figure}

In the downlink, it is assumed that the AP utilizes large number of antennas for beamforming. Moreover, if the modulation level used in the downlink is M-ary, then there are $M$ sets of weights, where for every symbol (i.e., class) we have a set of weights, as shown in Figure (\ref{Fig1}). 

It is assumed that there are $s$ antennas installed at the AP as shown in Figure (\ref{Fig1}). 
For class $i$, there is a set of weights,
\begin{equation}
\textbf{w}_i=\left[w_{i1},w_{i1},\dots, w_{is}\right]^H
\end{equation} 

where $H$ refers to the Hermitian transpose. Using the proper weight vector, one may control the relative phase and magnitude of the signal at each transmit antenna. The objective is to create a pattern of constructive and destructive interference in the wave-front at certain locations (terminals' locations). At a given time instance, all terminals in the downlink are classified within classes $i =1,\dots,M$, where every terminal can be in one and only one class corresponding to its received symbol at that time instance.  Assume all active terminals belong to set $U=\{u_1,u_2,\dots,u_N\}$. The weights of every class should be adjusted to achieve the following criteria for all $i=1,2,\dots M$ \cite{Elmu}

\begin{eqnarray}
\textbf{w}_i^H\textbf{a}_n=1, \,  \forall n \in  I=Class_i \subseteq U \\
\textbf{w}_i^H\textbf{a}_k=0, \, \forall k\in K; K\cap I=\emptyset; K\cup I= U
\end{eqnarray}

Where $\textbf{a}_n=\left[1, exp\left(j \triangle \Psi_{n2}\right), exp\left(j \triangle \Psi_{n3}\right),\dots exp\left(j \triangle \Psi_{ns}\right)\right]^H$ is the steering vector toward terminal $n$, and $\triangle \Psi_{nx}$ is the relative phase difference of the signal from antenna $x$ to the terminal $n$ and its value depends on the configuration of the transmit antenna array and the relative location of the terminal.  

The optimum weight values for class $i$ can be proved to be \cite{Godara}

\begin{equation} \label{Eqq}
\textbf{w}_i^H=\textbf{D}_i\left(\textbf{A}^H\textbf{A}\right)^{-1}\textbf{A}^H,\, i=1,2,\dots M
\end{equation}

Where $\textbf{D}_i = \left[0,1,0,\dots, 1,0,\dots\right]; 1's$ are located for terminals $n\in I$ and $0's$ for terminals $k \in K$, and $\textbf{A}=\left[\textbf{a}_1, \dots, \textbf{a}_N\right]$.  
In case of binary system, there will be just two classes and only two sets of weights.  Here we should emphasize that Null-Steering algorithm has been just used for demonstration of the proposed algorithm. It does not mean that it is the best algorithm to be used for this purpose. Two inherent difficulties with null-steering algorithm may reduce its attraction for practical application. One is the need to make inverse of a matrix with size equal to the number of terminals. When the number of terminals is large, we may have accumulation of numerical errors due to the limited accuracy of computers. Second problem is the need for the steering vector of all terminals (and their paths, in multipath scenario). 

\section{Relation with Shannon Capacity: Discussion}
Is this new BBMA algorithm of allocating resources over symbols rather than terminals violating Shannon upper capacity? Shannon capacity sets the upper bound of data rate in point to point communication due to noise at the receiver side. In that sense, Shannon capacity is always valid. However, the capacity is derived based on abstract treatment of information theory. The theory ignores the fact that in digital communication, information are formulated through a finite set of symbols. Therefore, similar symbols can assist each other than being interfering with each other. Furthermore, based on the Shannon capacity, it has been believed for long  that the capacity is always limited by the maximum available transmit power budget.  BBMA algorithm shows that this results is not necessary true. We will discuss this for downlink multi-user environment and also for a single link point-to-point scenario. For multi-user environment, by revisiting Equation (\ref{eq1}), we can formulate the total ergodic capacity such as

\begin{eqnarray} \label{shan}
C=\sum_{i=1}^N E_{G_i}\left[log_2\left(1+\frac{P_i G_i}{G_i\sum_{j=1, j\ne i}^N P_j \alpha_{ij}+\sigma_n^2}\right)\right]  \\ 
\sum_{i=1}^N P_i\left(t\right) \leq P_{max} \nonumber
\end{eqnarray}    

From Equation (\ref{shan}), it is clear that the total capacity of the downlink is limited (beside the noise) by the power budget at the AP. However, since by allocating the power to the symbols and divide the terminals in classes, we could support infinite number of terminals. In traditional communication, even in the presence of infinite number of parallel channels, the total ergodic capacity will be finite and limited by the available power budget. However, with this BBMA, the capacity in the sense of number of terminals can be infinite with finite power budget.

\begin{figure}[h]
\caption{BBMA for Point-to-Point Communication}
\label{Sha}
\centering
\includegraphics[width=1\textwidth]{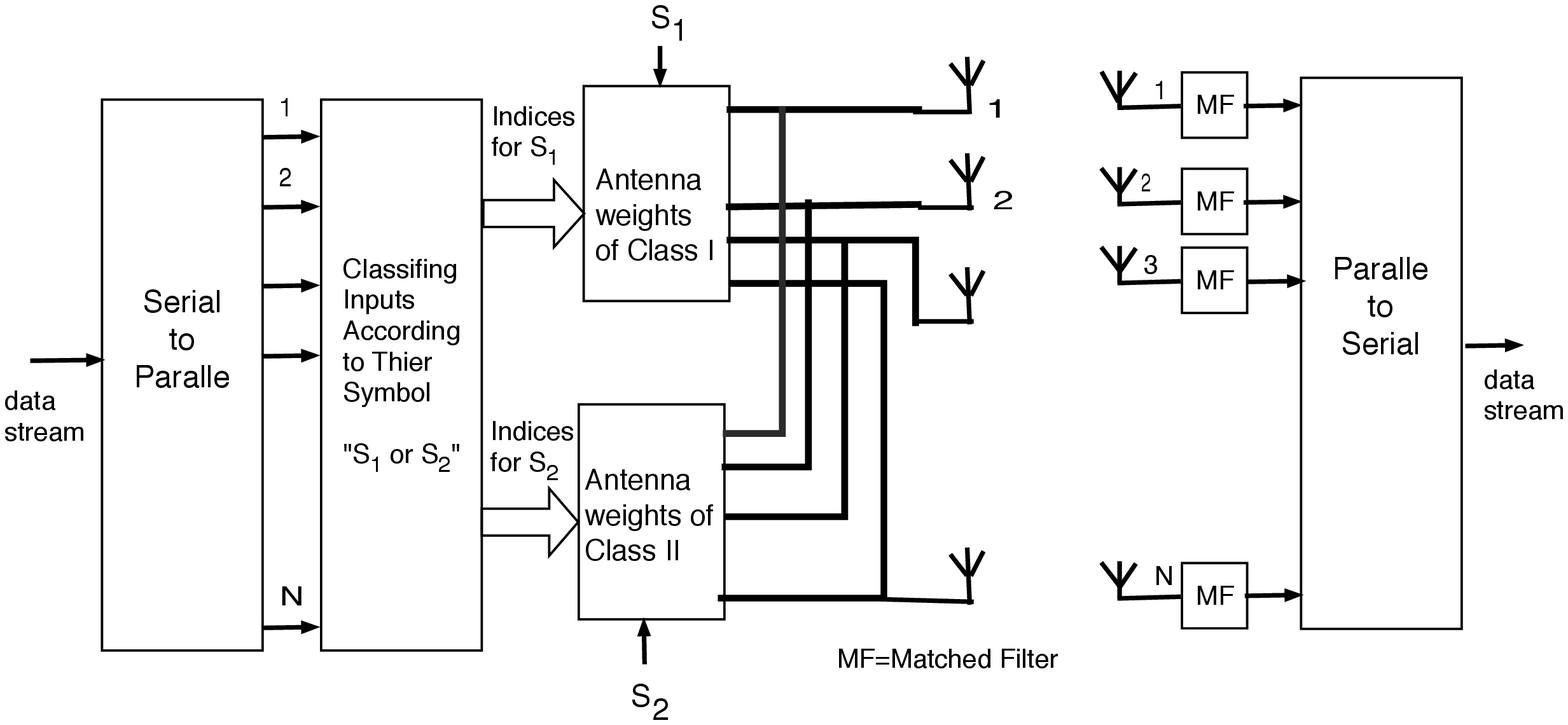}
\end{figure}

Moreover, it is possible to reconfigure the downlink of point-to-multipoint to single link point-to-point communication with infinite data rate. Figure (\ref{Sha}) shows the proposal to achieve very high data rate communication link. The concept is based mainly on using the same concept of BBMA but with single terminal and $N$ number of antennas. The number of antennas at the receiver should be exactly the same number of transmitted bits (or symbols). For example, with 256-ary system, then $N=8$. Furthermore, the number of the antennas at the transmitter should be at least the same number of the receiver antenna or more. This is required to guarantee the orthogonality between both classes. There are only two classes, class for bits with symbol $S_1=0$ and class for bits with symbol $S_2=1$ (multi-level symbols can be used as well). For example, if the transmitted word is $11010100$, then the bits' indices $"3,5,7,8"$ are assigned to  $Class_1$ and bits' indices $"1,2,4,6"$ are assigned to $Class_2$. Notice that the bit index is the same as the antenna number at the receiver terminal. Based on BBMA algorithm discussed before, it is possible to divide the power between those two classes in the transmitter side. To achieve the target signal quality in terms of probability of error, it is possible to specify the required transmit power per symbol (or symbol energy) \cite{porakis}. The total required energy does not depend on the number of bits. With orthogonality achieved by spatial processing, it does not matter if the transmitted bit is one or hundreds (as far as we have the required number of antennas at both sides). This again seems like violating Shannon capacity. Based on Shannon capacity algorithm, it was believed that infinite data rate will need infinite power budget (when the noise is not zero as it is always). However, with this BBMA algorithm, the limit is not the transmit power budget anymore, but the processing complexity (like how massive MIMO processing).  

\section{Simulation Results}

\begin{figure}
\caption{Transmit power with number of terminals}
\label{Fig_sim}
\centering
\includegraphics[width=1\textwidth]{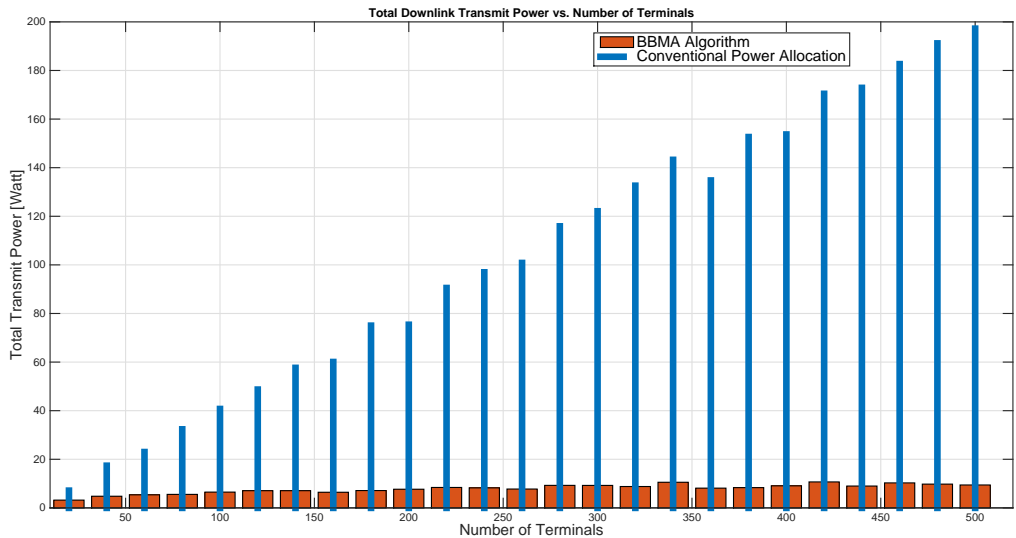}
\end{figure}

In this simulation, a single cell with an area of 1 $km^2$ and randomly distributed terminals is considered.  The AP uses  binary modulation such as BPSK in the downlink. The access point utilizes massive beamforming with 64 by 64 (two dimensional) antennas array.  Since we have just two symbols, then the system will utilize only two sets of weights. For every symbol, the AP would compute new weights set based on the distribution of terminals in each class.  Usually the time duration of symbols is very small compared to the changes in steering vectors, especially in low mobility terminals. This means simply the changing of the row vector $\textbf{D}$ only. The matrix $\textbf{A}$ in Equation (\ref{Eqq}) could be updated slowly such as once every several seconds or even at slower rate. However, this depends of course on the mobility nature of the terminals within the cell area. In the point-to-point configuration the matrix  $\textbf{A}$ is almost fixed.   The distance between antennas is $\frac{\lambda}{2}$ in both dimensions, where $\lambda$ is the carrier signal wavelength. In the conventional power allocation, we assumed full orthogonality between terminals which is achieved by the massive array (beamforming) at the transmitter. The target SINR has been set to $15 dB$ and all terminals have the same noise figure value of $10 dB$. There is also log-normal shadow fading with 8 dB standard deviation. The system bandwidth is 10 MHz. For the conventional power allocation we assumed perfect power control so that the power at the AP is set to guarantee the target SINR at each terminal input \cite{Elmu}. 
Figure (\ref{Fig_sim}) shows a comparison between the average required transmit power in two cases: the proposed technique and the conventional power allocation method.  It is clear that for traditional power allocation for terminals that even with perfect orthogonality, the required power increases linearly  with the number of terminals without a limit. However, in case of the BBMA method the transmit power is almost constant regardless of the number of terminals.

\begin{figure}
\caption{BER due to the Numerical Errors of the Inversion of Matrix \textbf{A}}
\label{fig2}
\centering
\includegraphics[width=1\textwidth]{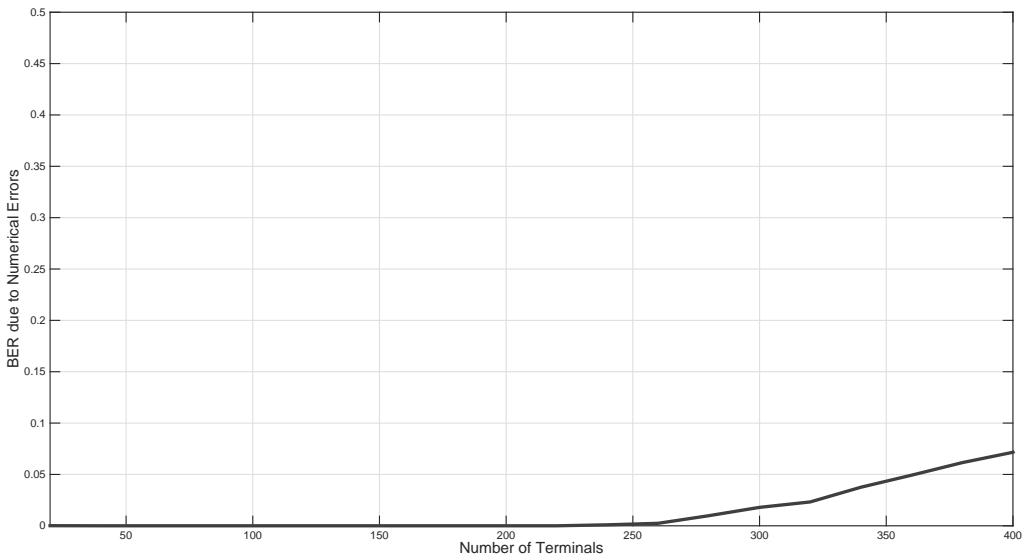}
\end{figure}

Actually, this is one of the most attractive features of the proposed technology that, at least theoretically, it can support very large terminals with a limited power budget. 
However, since the null-steering beamforming algorithms depends on solving large system of linear equations. The numerical errors increase considerably when we have ill-conditioned matrix $\textbf{A}$. The average BER for the simulation is shown in Figure (\ref{fig2}).  Based on the simulation results, the numerical errors start to appear when the number of terminals exceeds $200$. Therefore, more reliable beamforming algorithm with higher performance and robustness should be used. 

\begin{figure}
\caption{Transmit power with number of terminals}
\label{sha_qsk}
\centering
\includegraphics[width=1\textwidth]{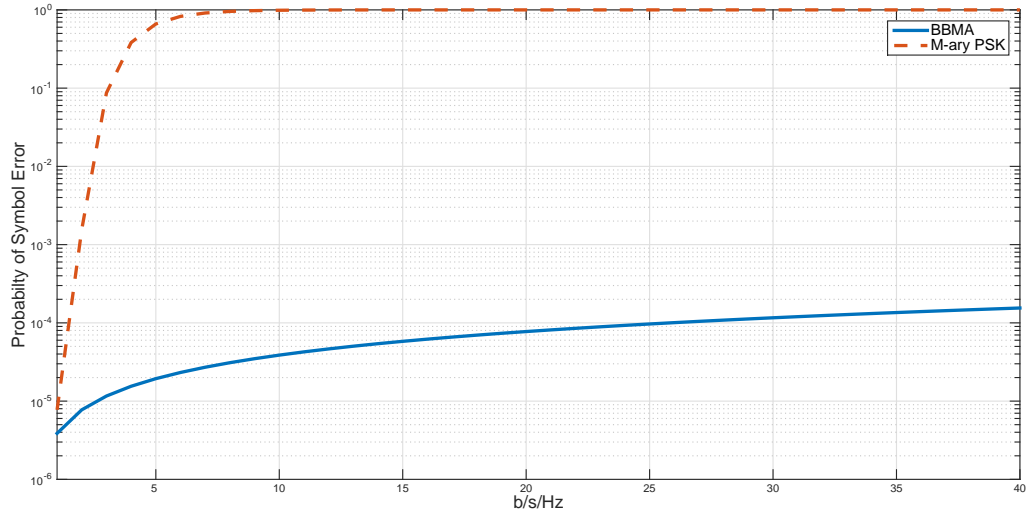}
\end{figure}

It has been shown that the same concept of allocating power directly to symbols could be extended to the case of point-to-point communication. Since the same power needed for a single symbol "bit" is used for all symbols after achieving the orthogonality by spatial processing. Therefore,  the energy efficiency of using BBMA can be set as high as desirable without any theoretical limit. However, the technical limit is only the complexity of the spatial processing. For point-to-point communication, assume at the receiver side a fixed symbol energy $E_s=1 J$, and a constant $\frac{E_s}{N_0}=10$. The probability of symbol error versus the spectral efficiency in  terms of $b/s/Hz$ is shown in Figure (\ref{sha_qsk}). It is shown for two cases, one with using M-ary PSK modulation (SISO) and with the proposed algorithm of BBMA. The M-ary PSK is given just as a reference. We this configuration, approaching infinite data rate is achieved with approaching zero Joule per bit, but with approaching infinite complexity in terms of spatial processing.  

\section{Conclusions}

New concept for power allocation is introduced in this paper. The power is allocated per transmitted symbol rather than for each terminal. Therefore, information transmission is performed through broadcasting of symbols instead of the traditional way of dedicating power for each receiver. However, it is needed to classify between receiver due to their received symbol at each symbol duration. This could be achieved by using spatial processing, e.g. null-steering. It has been shown that, for the downlink in cellular networks, the required power does not depend on the number of supported terminals. Theoretically, even infinite number of terminals could be supported at finite limited power budget. The only limitations are the number of antennas and the numerical errors of matrix inversion to find the optimum antenna weights. This could be very useful for green communication systems. This could also considerably increase the network capacity due to reduction of interference levels. New configuration for point-to-point communication has been also proposed. It has been shown that with finite power budget, it is possible to have any data rate with small probability of error. The only limitation is the number of antennas at the spatial processing. However, this opens interesting question, does BBMA algorithm violates Shannon capacity limit? Some technical discussions have been given, however, the topic will need more deep analysis in terms of information theory. The main target of this paper is to introduce this new multi-content broadcasting algorithm and show its features.

\section*{Acknowledgments}
The author would like to acknowledge the suggestions and deep discussions with Dr. Naser Tarhuni from Qaboos University in Oman and Prof. Riku Jantti from Aalto University in Finland. 

\nocite{*}
\bibliographystyle{IEEE}

%

%

\end{document}